\documentclass{PoS}

\usepackage{epsfig}
\usepackage{cite}

\def\({\left(}
\def\){\right)}
\def\[{\left[}
\def\]{\right]}

\relax

\def    \hepph  #1 {{\tt hep-ph/#1}}
\def    \hepex  #1 {{\tt hep-ex/#1}}

\def\Black{}

\newsavebox\tmpfig

\newcommand{\lc}{\left[}
\newcommand{\rc}{\right]}
\newcommand{\lp}{\left(}
\newcommand{\rp}{\right)}
\newcommand{\be}{\begin{equation}}
\newcommand{\ee}{\end{equation}}

\title{Progress in the Neural Network Determination 
of Polarized Parton Distributions}

\ShortTitle{Polarized NNPDFs}

\author{\speaker{J. Rojo}, S. Forte\\
         Dipartimento di Fisica, Universit\`a di Milano and
INFN, Sezione di Milano,\\ Via Celoria 16, I-20133 Milano, Italy\\
        E-mail: \email{juan.rojo@mi.infn.it}, \email{stefano.forte@mi.infn.it}
}

\author{G. Ridolfi\\
        INFN Sezione di Genoa\\
        E-mail: \email{ridolfi@ge.infn.it}
}

\author{R. D. Ball. L. Del Debbio, M. Ubiali\\
         School of Physics and Astronomy, University of Edinburgh,\\
JCMB, KB, Mayfield Rd, Edinburgh EH9 3JZ, Scotland\\
        E-mail: \email{rdb@ph.ed.ac.uk}, 
\email{luigi.del.debbio@ed.ac.uk}, \email{maria.ubiali@gmail.com}
}

\author{V. Bertone, A. Guffanti\\
         Physikalisches Institut, Albert-Ludwigs-Universit\"at Freiburg
\\ Hermann-Herder-Stra\ss e 3, D-79104 Freiburg i. B., Germany \\
        E-mail: \email{Valerio.Bertone@physik.uni-freiburg.de}, 
\email{alberto.guffanti@physik.uni-freiburg.de}
}

\author{F. Cerutti, J. I. Latorre\\
        Departament d'Estructura i Constituents de la Mat\`eria, 
Universitat de Barcelona,\\ Diagonal 647, E-08028 Barcelona, Spain\\
        E-mail: \email{francesco.cerutti@gmail.com}, \email{latorre@ecm.ub.es}
}

\abstract{We review recent progress 
towards a determination of a set of polarized parton distributions
 from a global set of 
deep-inelastic scattering data based on the NNPDF methodology,
in analogy with the unpolarized case. This method
is designed to provide a faithful and
statistically sound representation of parton distributions and their 
uncertainties. We show how the
FastKernel method provides a fast
and accurate method for solving the polarized
DGLAP equations. 
We discuss the polarized PDF parametrizations and the
physical constraints which can be imposed. Preliminary
results suggest that the uncertainty on
polarized PDFs, most notably the gluon, has been underestimated
in previous studies.}

\FullConference{DIS 2010}

\begin{document}

\paragraph{Polarized PDFs with the NNPDF approach}

The interest in polarized deep inelastic scattering was revived in 1988 
by the results of the
EMC experiment that led to the so-called ``spin crisis''. Since then a
lot of progress has been made~\cite{Altarelli:1998nb,Kuhn:2008sy,Vogelsang:2004hg}. Several experiments have been completed at CERN, 
SLAC, DESY and JLAB, with several more are ongoing including RHIC, the first
polarized hadronic collider, and on the theory side, it was understood that the
``spin crisis'' is a sign of the non-trivial spin
structure of the nucleon, which should be
understood in terms of QCD. While on a first stage interest was focused
on the determination of the first moments of the polarized parton densities
and the associated polarized sum rules, in the recent years
attention is now shifting to the full reconstruction of polarized parton 
densities, particularly the gluon density. The current bottleneck
is the accurate determination of the uncertainties on polarized
parton distributions, a problem which in the unpolarized case
is starting to be solved only in recent times.

On the other hand, one of the most important advances in unpolarized global PDF
analysis in the recent years has been the developement
of the NNPDF
methodology~\cite{DelDebbio:2007ee,Ball:2008by,Ball:2009mk,Ball:2009qv,Ball:2010de}. NNPDF provides a 
determination  of unpolarized 
PDFs and their uncertainty which is  independent of the
choice of data set, and which has been shown in benchmark
studies~\cite{Dittmar:2009ii} to behave in a statistically consistent
way when data are added or removed to the fit. The use
 of artificial neural networks as unbiased interpolatants is crucial to obtain unbiased results which
are independent of the choice of input functional form for the
PDFs, which is specially relevant for those PDF combinations which
are loosely constrained by data. Also, because of the use of a
Monte Carlo approach, the NNPDF methodology is easily amenable 
to the use of standard
statistical tools, and does not rely on any of the usual gaussian
approximations used for the PDF uncertainty estimation and
determination in many analysis.

In this contribution we review progress towards the
applications of the NNPDF approach to the determination of a
polarized structure functions based on inclusive polarized
DIS data: NNPDFpol1.0.\footnote{Note that the NNPDF approach was applied
to the determination of polarized asymmetries from inclusive
polarized data in Ref.~\cite{DelDebbio:2009sq}} We show that once the bias
from the choice of fixed functional forms are removed, the uncertainty on
some polarized PDFs, most notably the gluon, are rather larger than
previously estimated.

\paragraph{The NNPDFpol1.0 analysis}

\begin{figure}[ht]
\begin{center}
\epsfig{width=0.60\textwidth,figure=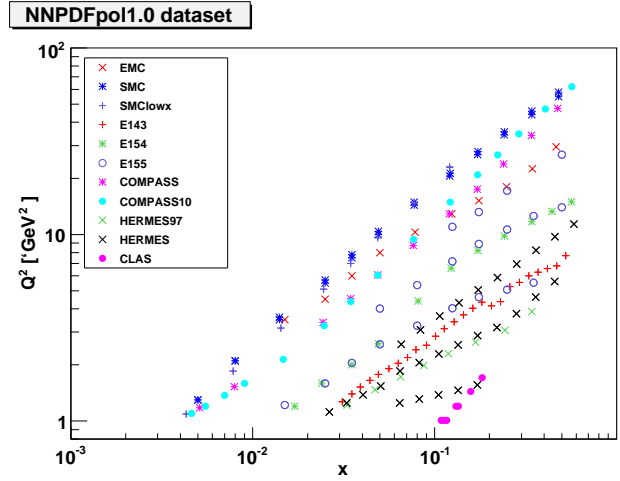}
\caption{\small
\label{fig:kin} The kinematical coverage of the
inclusive polarized DIS experiments included in the
NNPDFpol1.0 analysis.} 
\end{center}
\end{figure}

The first NNPDF polarized analysis will be NNPDFpol1.0.
Fig.~\ref{fig:kin} shows the inclusive polarized DIS experiments 
and their kinematical coverage included in the
NNPDFpol1.0 analysis. The number of data points after kinematical
 cuts is $N_{\rm dat}\sim 250$,
about an order of magnitude smaller than in the unpolarized case.
The kinematical cuts applied $Q^2\ge 1$ GeV$^2$ and 
$W^2\ge 6.25$ GeV$^2$ find a compromise between keeping the maximum
number of data where remaining in the perturbative region and
removing dynamical higher twist effects\cite{Simolo:2006iw}.

%%%%%%%%%%%%%%%%%%%%%%%%%%%%%%%%%%%%
%%%%%%%%%%%%%%%%%%%%%%%%%%%%%%%%%%%%%%

As in the unpolarized case, the polarized PDF evolution as
implemented in the FastKernel framework has been benchmarked
with the Les Houches benchmark tables of Ref.~\cite{Dittmar:2005ed}, 
obtained
from the HOPPET~\cite{Salam:2008qg} and PEGASUS~\cite{Vogt:2004ns} 
evolution packages. Results of this
benchmark comparison are shown in Table~\ref{tab:lhacc}, where
it can be seen that the accuracy is excellent for all
data points and all polarized PDFs.

%%%%%%%%%%%%%%%%%%%%%%%%%%%%%%%%%%
\begin{table}[ht]
\begin{center}
\vskip-0.1cm
\small
\begin{tabular}{|c||c|c|c|c|}
\hline
$x$  &  $\epsilon_{\rm rel}\lp \Delta u_V\rp$ & 
 $\epsilon_{\rm rel}\lp \Delta d_V\rp$ &  $\epsilon_{\rm rel}\lp \Delta \Sigma\rp$ &
 $\epsilon_{\rm rel}\lp   \Delta g\rp$  \\
\hline
\hline   
$10^{-3}$ & $1.1\,10^{-4}$ & $9.2\,10^{-5}$ & $9.9\,10^{-5}$& $1.1\,10^{-4}$\\
$10^{-2}$  & $1.4\,10^{-4}$ & $1.9\,10^{-4}$ & $3.5\,10^{-4}$& $9.3\,10^{-5}$\\
$0.1$  & $1.2\,10^{-4}$ & $1.6\,10^{-4}$ & $5.4\,10^{-6}$& $1.7\,10^{-4}$\\
$0.3$  & $2.3\,10^{-6}$ & $1.1\,10^{-5}$ & $7.5\,10^{-6}$& $1.7\,10^{-5}$\\
$0.5$  & $5.6\,10^{-6}$ & $9.6\,10^{-6}$ & $1.6\,10^{-5}$& $2.5\,10^{-5}$\\
$0.7$  & $1.2\,10^{-4}$ & $9.2\,10^{-7}$ & $1.6\,10^{-4}$& $7.8\,10^{-5}$\\
$0.9$  & $3.5\,10^{-3}$ & $1.1\,10^{-2}$ & $4.1\,10^{-3}$& $7.8\,10^{-3}$\\
\hline
\end{tabular}
\end{center}
\caption{\small Comparison of the accuracy of our polarized PDF evolution
with respect to the Les Houches benchmark tables for different
polarized PDF combinations at NLO in the ZM-VFNS. 
\label{tab:lhacc}}
\vskip-0.1cm
\end{table}.
%%%%%%%%%%%%%%%%%%%%%%%%%%%%%%%%%

In this analysis four polarized PDFs are parametrized
with artificial neural networks. The specific basis which
 we choose at the initial evolution scale $Q_0^2=1$ GeV$^2$ is given by the
following linear combinations:
 \begin{itemize}
\item the singlet distribution, $\Delta\Sigma(x)\equiv \sum_{i=1}^{n_f}\lp
\Delta q_i(x)+\Delta \bar{q}_i(x)\rp$, 
\item the non-singlet triplet, $\Delta T_3(x) \equiv 
\lp \Delta u(x)+ \Delta\bar{u}(x)\rp - \lp 
\Delta d(x)+ \Delta \bar{d}(x)\rp$, 
\item the non-singlet octet, $\Delta T_8(x) \equiv 
\lp \Delta u(x)+ \Delta\bar{u}(x)\rp + \lp 
\Delta d(x)+ \Delta \bar{d}(x)\rp - 2 \lp 
\Delta s(x)+ \Delta \bar{s}(x)\rp$, 
\item the gluon, $\Delta g(x)$ \ .
\end{itemize}
Each of these polarized PDFs has 37 free parameters (2-5-3-1 architecture)
to be determined from experimental data using the minimization
strategy discussed in Ref~\cite{Ball:2008by}. Heavy quark PDFs are generated
dynamically, and heavy quark mass effects can be taken
into account using the FONLL general-mass scheme~\cite{Forte:2010ta}.

An important constraint on the 
normalization of the polarized triplet and octet can be provided
by the axial sum rules~\cite{Kuhn:2008sy},
\be
\label{eq:t3sr}
\lc \Delta T_3(Q_0^2) \rc \equiv \int_0^1 dx~ \Delta T_3(x,Q_0^2)=a_3 \ ,
\ee
\be
\label{eq:t8sr}
\lc \Delta T_8(Q_0^2) \rc \equiv \int_0^1 dx~ \Delta T_8(x,Q_0^2) =a_8 \ ,
\ee
where $a_3$ and $a_8$ are respectively the triplet and octet
axial charges, which can be determined from weak baryon decays,
\begin{equation}
\label{eq:axial}
a_3=g_A=1.2670\pm0.0035\, \qquad a_8 = 0.585 \pm 0.025 \ , 
\end{equation}
The value of $a_8$ assumes exact SU(3) symmetry, the effects of 
potential SU(3) violations can be accounted for by adding
a suitable theoretical uncertainty.

In the context of polarized structure functions, 
 positivity implies bounds on the size of the polarized structure 
 functions $g_1^p$ and $g_1^d$  determined by the size of the 
corresponding unpolarized 
structure functions $F_1^p$ and $F_1^d$ . We impose these 
bounds on $g_1^p$ and $g_1^d$using consistently the unpolarized structure functions as determined  in Ref.~\cite{Del Debbio:2004qj}.

\begin{figure}[ht]
\begin{center}
\epsfig{width=0.47\textwidth,figure=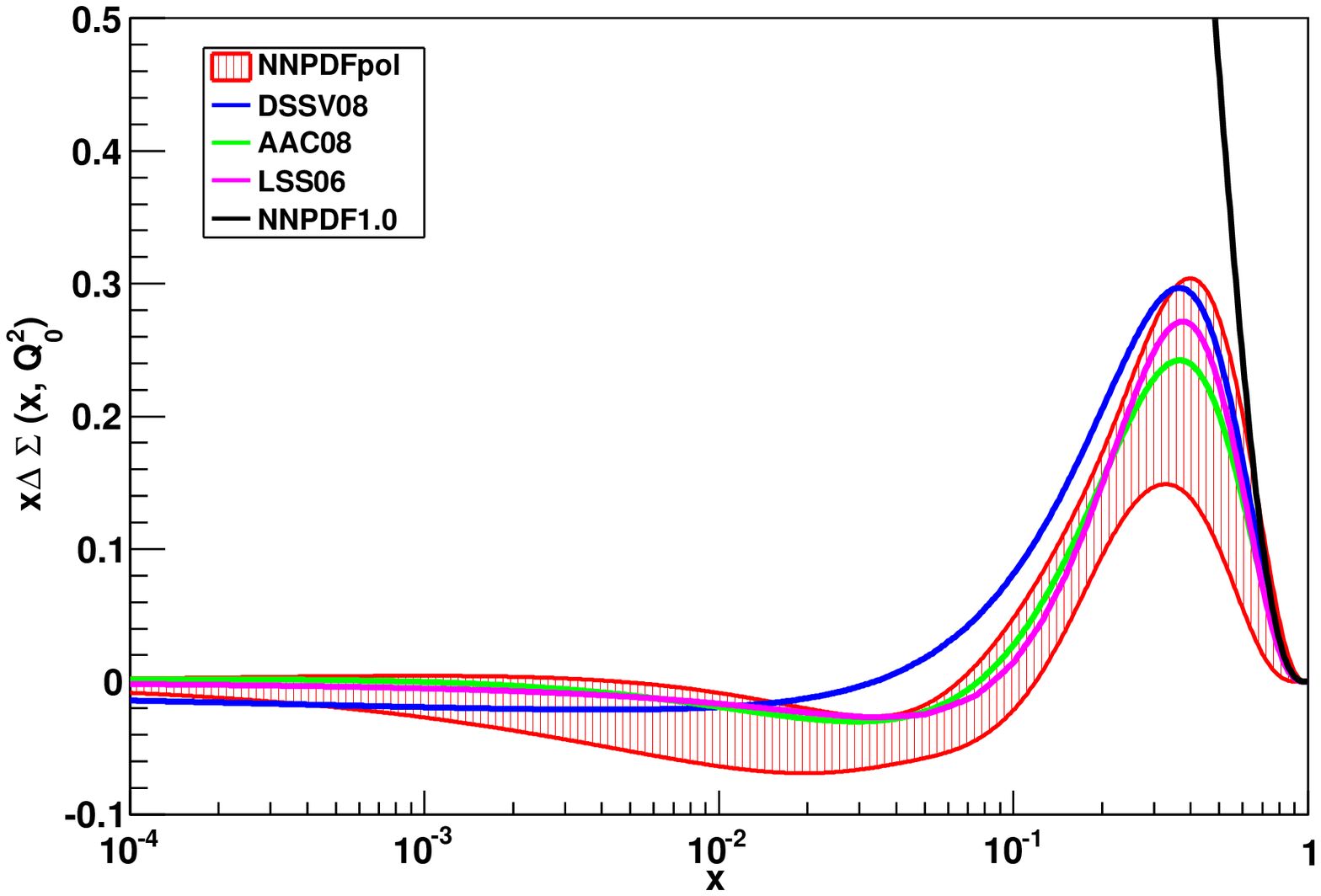}
\epsfig{width=0.47\textwidth,figure=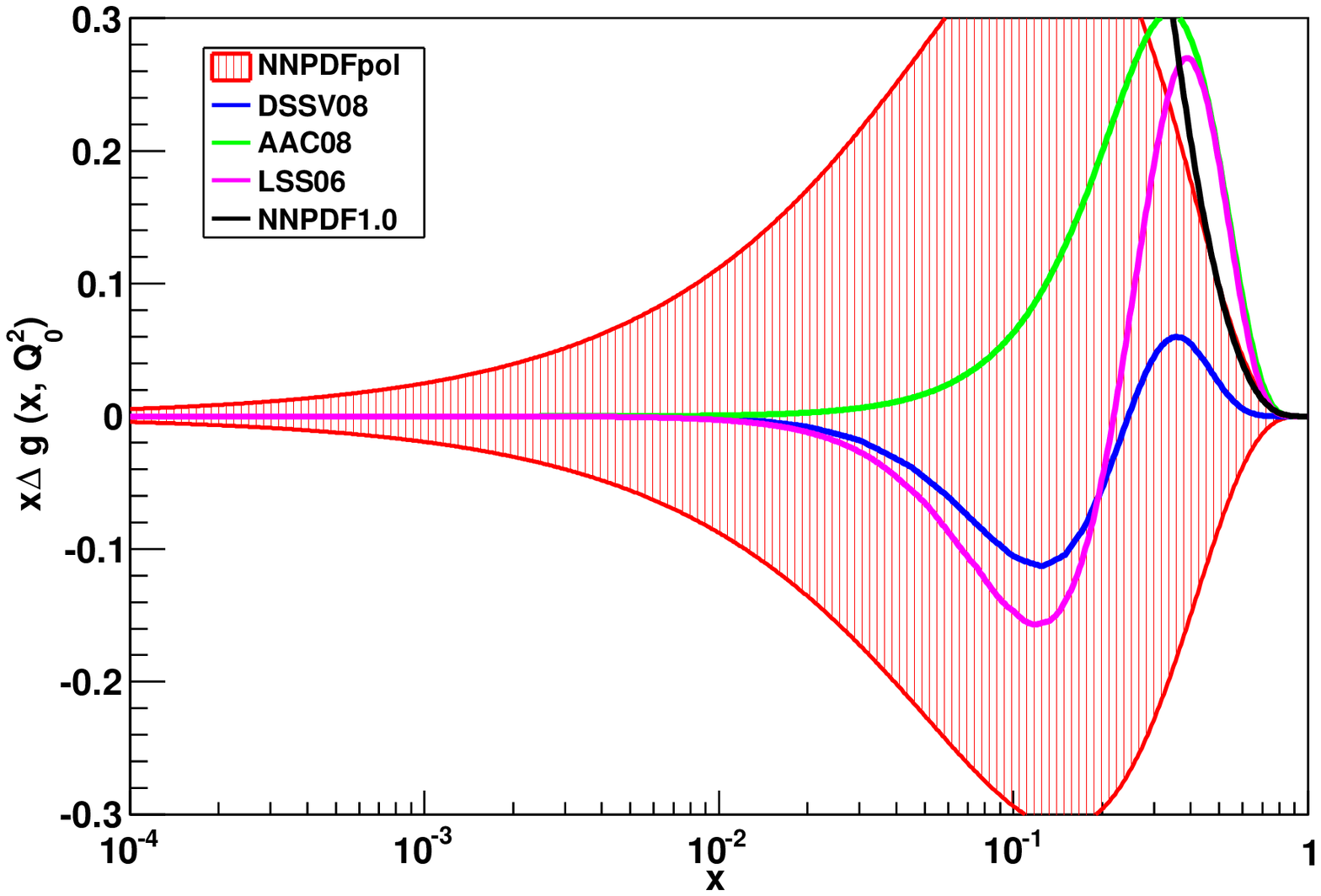}
\epsfig{width=0.47\textwidth,figure=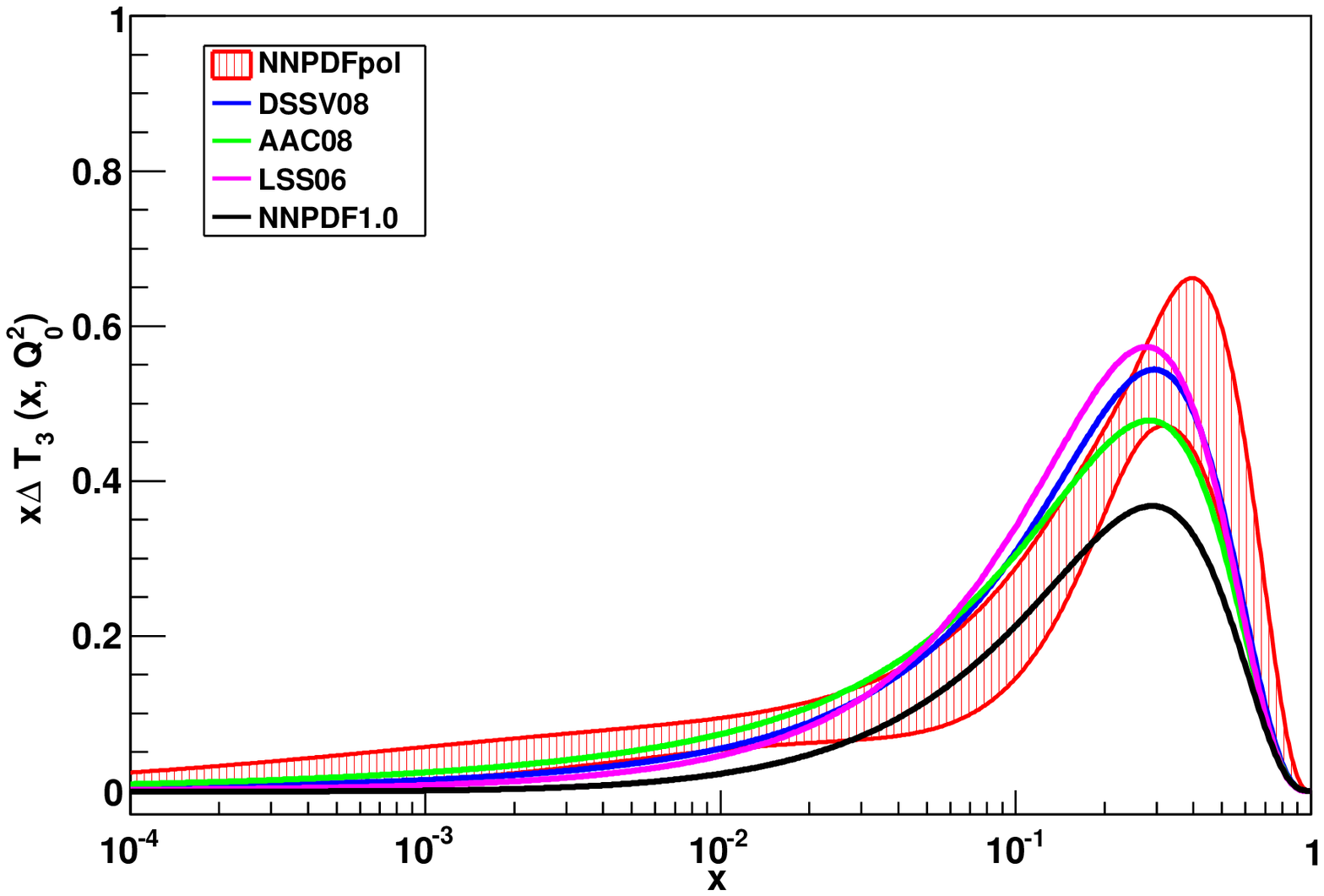}
\epsfig{width=0.47\textwidth,figure=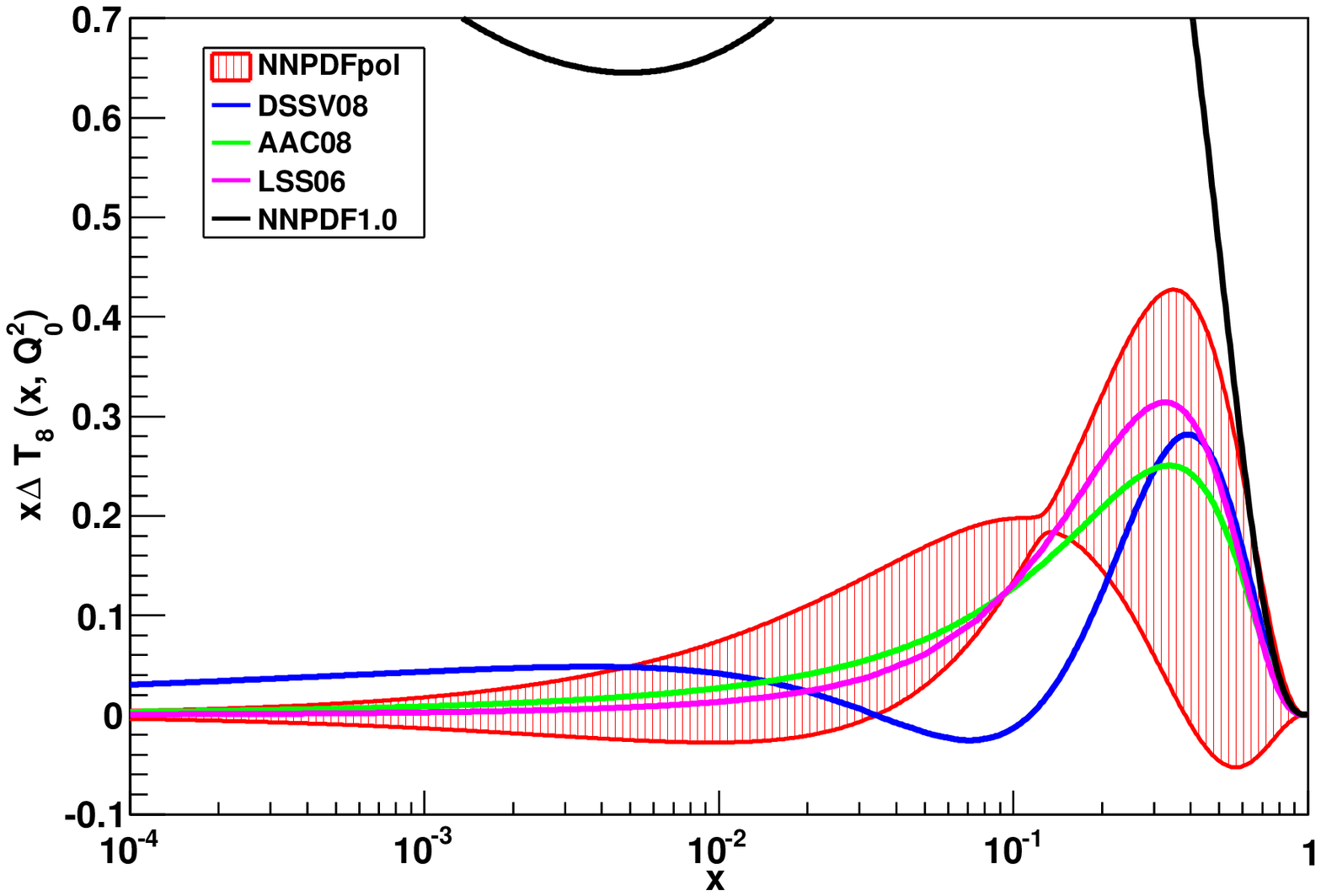}
\caption{\small
\label{fig:ppdfs} Comparison of the
NNPDFpol1.0 polarized parton distributions at the
initial evolution scale $Q_0^2=1$ GeV$^2$, compared
to other recent determinations.
For illustration, we also show the
corresponding unpolarized PDF set
NNPDF1.0.} 
\end{center}
\end{figure}

We show preliminary results for the NNPDFpol1.0 polarized PDF set
in Fig.~\ref{fig:ppdfs}, where they are compared to other recent 
polarized PDF determinations~\cite{deFlorian:2009vb,Leader:2006xc}.
 Although these results are
too preliminary to draw quantitative conclusions, they
seem to indicate that the uncertainty on the polarized gluon
from inclusive data is rather larger than previously assumed, and
in particular its sign cannot be determined.

%%%%%%%%%%%%%%%%%%%%%%%%%%%%%%%%%%%%
%%%%%%%%%%%%%%%%%%%%%%%%%%%%%%%%%%%%%%

\paragraph{Outlook}

In this contribution we have outlined recent progress
towards the generalization of the NNPDF methodology to the
polarized sector. We have presented preliminary results
for NNPDFpol1.0, a set of polarized PDFs obtained from
a global analysis of inclusive polarized DIS data
using the NNPDF approach.
NNPDFpol1.0 will be the first polarized PDF set which
is determined consistently together with its unpolarized
counterpart.
Eventually we hope to also include exclusive DIS data 
 and polarized hadronic data without any K--factor approximations using 
 the FastKernel method, as was done recently in our global fits to
 unpolarized data~\cite{Ball:2010de}.

\end{document}